\documentclass[12pt,a4paper]{article}
 \usepackage{times}
    \usepackage[english]{babel}
    \usepackage{amsmath,amssymb,amsfonts}
    \usepackage{textcomp}
    \usepackage{graphicx}
\begin{document}
\title{Fifth Memoir\\or\\
Letter on the resistance of air to the movement of pendulums\footnote{Originally published as {\it Cinqui\`eme M\'emoire ou Lettre sur la R\'esitance de L'Air ou Mouvement des Pendules}
in {\it M\'emoires sur diff\'erens sujets de Math\'ematiques}. Published in Paris by Pissot and Durand, 1748. This translation
has been prepared by S.R. Dahmen, Instituto de F\'{\i}sica, UFRGS, 91501-970 Porto Alegre, Brazil and
K.M. Pacheco, Facult\'e de Philosophie de l'Universit\'e de Strasbourg 7, 67000 Strasbourg, France. Revised by N. Maillard (Porto Alegre) and P. Mac Carron (Oxford).}.}
\author{Denis Diderot\footnote{Born: 5 October 1713 in Langres, France. Died: July 31 1784 in Paris, France.}}
\date{}
\maketitle{}

M***

If the place where Newton calculates the resistance caused by air to the movement of the pendulum embarasses you,
do not let your self-steem be afflicted by it. As the greatest geometers will tell you, in the depth and laconicity of the
Principia one encounters everywhere motives to console a man of penetrating mind who has had some difficulty in understanding them;
and you will see shortly that there is another reason that seems even better to me -- that the hypothesis this author started with might not be exact.
Something surprises me however: that you were
advised to seek me in order to free you from your embarassment. It is true that I studied Newton with the purpose of elucidating him. I should
even tell you that this work was pursued, if not sucessfully, at least with great vivacity. But I did not think of it any longer since the Reverend Fathers
Le Sueur and Jacquier made their commentaries public, and I did not feel tempted to ever reconsider it. There was, in my work, a few things you would not
find in the work of these great geometers and a great many things in theirs you most surely would not find in mine. So what do you ask of me?
Even though mathematical matters were once much familiar to me, to ask me now about Newton is to talk of a dream of the year past. However, to persevere
in the habit of pleasing you I will leaf through my abandoned drafts, I will consult the sagacity of my friends and tell you what I can learn
from them, telling you also, with Horace: 
\begin{quote}
 {\it `Si quid novisti rectius istis, candidus imperti; si non, hi utere mecum.'}~\footnote{Horace, Epistolae,
Liber I, ep. VI, vers. 67, 68 Edition. [Transl.: If you can better these, please tell me. If not, follow them
with me.]}
\end{quote}
\begin{center}
 {\bf Proposition I.\\Problem.}
\end{center}
\noindent
{\it Let a pendulum $M$, which describes an arc $BA$ in air,  be attached to a fixed point $G$ through the straight line $GM$. One asks for its
velocity at any given point $M$, given that it was let go from point $B$ (see fig. 2).}\\

\begin{figure}[h]
\centering
 \includegraphics[scale=0.5]{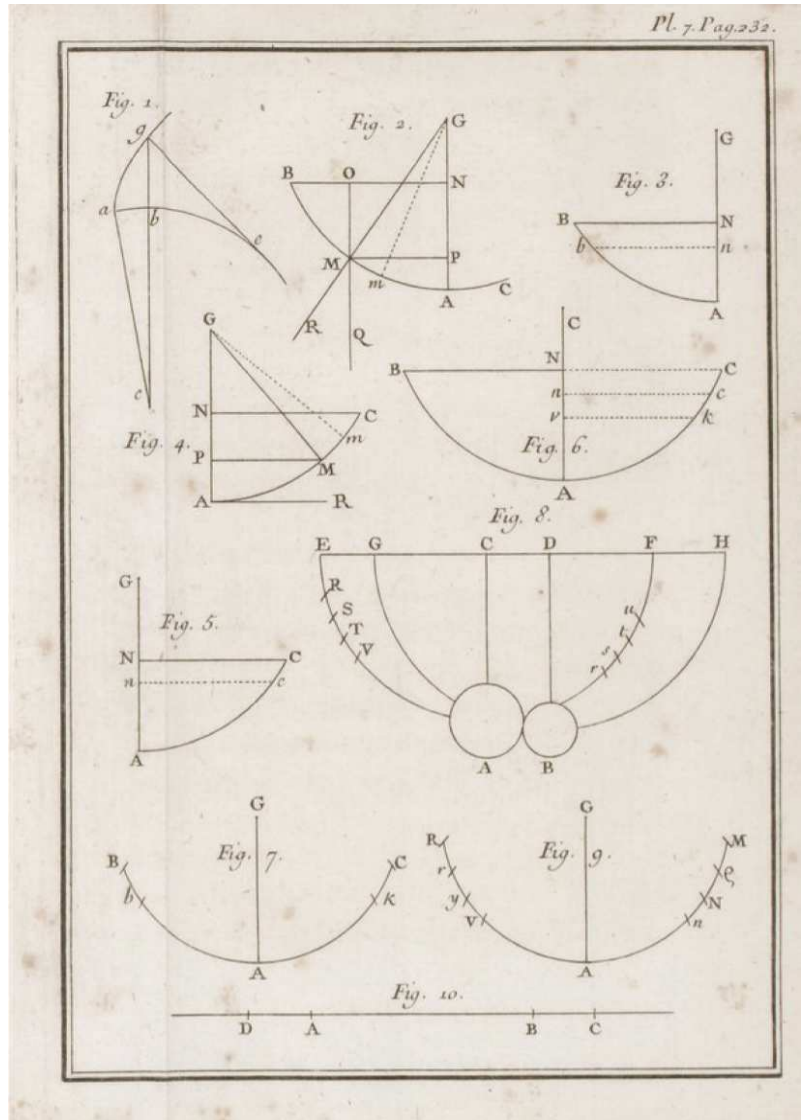}
 \caption{Figure from the original edition of the fifth memoir of 1748.}
\end{figure}
  \noindent
Let $GM=a$, $NA=b$, $AP=x$. The weight is $p$. The resistance that air causes on $M$ when it has a velocity $g$ is equal to $f$. The velocity
of the pendulum at point $M$ is equal to equal to $v$.
\vskip 0.5cm
\noindent
If one assumes, as all physicists do, that the resistance in air and in other fluids is proportional to the square of the velocity, the resistance
at point $M$ is equal to$fv^2/g^2$, and this resistance, acting along [the arc] $Mm$ will act so as to decrease the velocity. Moreover, one can easily
see that the weight $p$ acting  along $MQ$ can be decomposed into two other forces: one, acting along $MR$, is compensated by the resistance
of the string or straight line $GM$, while the other acts along $Mm$, perpendicularly to $GM$, and is equal to
\begin{equation}
 \frac{p\times\; MP}{GM}=\frac{p\;\sqrt{2ax - x^2}}{a}.\nonumber
\end{equation}
So, the total accelerating force at point $M$ that causes the body to move along $Mm$ is equal to
\begin{equation}
\frac{p\;\sqrt{2ax - x^2}}{a} - \frac{fv^2}{g^2}.\nonumber
\end{equation}
But the time it takes to traverse $Mm$ is equal to $Mm/v$ and the element or increase in velocity is equal to tha accelerating force
multiplied by time. Thus~\footnote{In the original there is a mass $m$ missing on the right hand side of the equation. So, for the following
equations to hold, one has to consider $m=1$. [N. of T.]}
\begin{equation}
\biggl(\frac{p\;\sqrt{2ax - x^2}}{a} - \frac{fv^2}{g^2}\biggr)\times\frac{Mm}{v}= dv.\nonumber
\end{equation}
In this equation I substitute the little arc
$Mm$ by its value $-\frac{a\;dx}{\sqrt{2ax - x^2}}$, with a minus sign, because as the pendulum descends
the velocity increases while x becomes smaller. I have
\begin{equation}
p\;dx  + \frac{fv^2\times adx}{g^2\sqrt{2ax - x^2}}=v\;dv.\nonumber
\end{equation}
whose integral is
\begin{equation}
 \frac{v^2}{2} = pb-px + \int\frac{f\;v^2\times\;a\;dx}{g^2\sqrt{2ax - x^2}}.\nonumber
\end{equation}
I added the constant $pb$ since $v=0$ when $x=b$, that is when the pendulum is at point $B$ it falls due to
its own weight.

First, one will notice in this equation that if a pendulum falls in vacuum or in a non-resistant medium, it will
have a velocity [given by]  $v^2 = pb - px$. However, as the resistance of air is much smaller than the weight $p$, the real value of
$v^2$ will differ little from $2pb - 2px$ and one may then replace $fv^2$ by $f(2pb - 2px)$, and this will cause
but a very small error.

Thus one has
\begin{equation}
v^2 = 2pb - 2px + 2 \int\frac{f(2pb-2px)\times\;adx}{g^2\sqrt{2ax - x^2}}.\nonumber
\end{equation}
for the approximate value of $v^2$.
It is now a question of finding the integral of the term under the sign $\int$, and the difficulty reduces to integrating
$\frac{ba\;dx - ax\;dx}{\sqrt{2ax - x^2}}$.
It should be remarked that this integral is such that it must be $0$ when $x=b$. Or, the integral of the first
term $\int \frac{ba\;dx}{\sqrt{2ax - x^2}}$ is $b\times\;(arc\; AM -\;arc\;AB)$. To this I added the constant $-b\times arc\;AB$
so that $\int \frac{ba\;dx}{\sqrt{2ax - x^2}}$ is equal to $0$ when x is equal to $b$; thus one has
\begin{equation}
 \int \frac{ba\;dx}{\sqrt{2ax - x^2}}= -b\times\;arc \; BM.\nonumber
\end{equation}
Now, to find the integral $\int \frac{-a\;dx}{\sqrt{2ax - x^2}}$, I write it as
\begin{equation}
 \int\frac{-a\;x\;dx}{\sqrt{2ax - x^2}} = \int\frac{a^2\;dx - ax\;dx}{\sqrt{2ax - x^2}} -\int\frac{a^2\;dx}{\sqrt{2ax - x^2}}.\nonumber
\end{equation}
where the integral is $a\sqrt{2ax-x^2} -a\times\; AM = a \times\;(MP-AM)$, to which the constant $-a (BN-AB)$ must be added
for the same reason explained above; one will have then
\begin{equation}
 \int\frac{-a\;x\;dx}{\sqrt{2ax - x^2}}= -a  \times\; (BO-BM).\nonumber
\end{equation}
Hence~\footnote{In the original, a factor $g^2$ is missing in the second term on the right hand side. [N. of T.]}
\begin{equation}
 v^2 = 2pb -2px -\frac{2f\times 2pb\times BM}{g^2}-\frac{2f\times 2pa\times (BO-BM)}{g^2}.\nonumber
\end{equation}
\vskip 0.5cm
\begin{center}
 {\bf Corollary I.}
\end{center}
When the pendulum arrives at $A$, one has
\begin{equation}
v^2 = 2pb -\frac{2f\times 2pb\times BA}{g^2}-\frac{2f\times 2pa\times (BN-BA)}{g^2}.\nonumber
\end{equation}
\vskip 0.5cm
\begin{center}
 {\bf Corollary II.}
\end{center}
Thus from (fig. 3), if one takes $An = b - \frac{2fb\times\;BA}{g^2}-\frac{2fa\times\;(BN-BA)}{g^2}$ one has
$v^2 = 2p\times\;An$, that is, the velocity at $A$ is the same as the one the pendulum would have
if it had fallen in vacuum from $b$ to $A$.
\vskip 0.5cm
\begin{center}
 {\bf Corollary III.}
\end{center}
If arc $AB$ has but a few degrees, $BN$ will be almost equal to $BA$; in this case one may assume
$v^2= 2pb-\frac{2f\times\;2pb\times\;BA}{g^2}$
\vskip 0.5cm

\begin{center}
 {\bf Proposition II.\\Problem.}
\end{center}
{\it Suppose (fig. 4) that a pendulum $A$, placed in the vertical $GA$, receives an impulse or velocity $h$ along the horizontal $AR$. 
One asks for its velocity at any given point $M$.}\\
\begin{center}
 {\bf Solution}
\end{center}
Using the same naming of variables as before, the retarding force will be 
\begin{equation}
\frac{p\;\sqrt{2ax - x^2}}{a} + \frac{fv^2}{g^2}.\nonumber
\end{equation}
as the resistance now helps the weight, continuously diminishing the velocity of the pendulum. Thus one will have~\footnote{There is a misprint in the original.
There appears a $du$ instead of a $dv$. [N. of T.]}
\begin{equation}
-dv=\biggl(\frac{adx}{v\;\sqrt{2ax - x^2}}\biggr)\times\biggl(\frac{p\;\sqrt{2ax - x^2}}{a} +\frac{fv^2}{g^2}\biggr).\nonumber
\end{equation}
I write $-dv$ since, as $x$ increases, $v$ decreases, so that
\begin{equation}
-vdv= pdx + \frac{fv^2\times\; adx}{g^2\;\sqrt{2ax - x^2}}.\nonumber
\end{equation}
and adding the constants
\begin{equation}
\frac{h^2-v^2}{2}= px +\int \frac{fv^2\times\; adx}{g^2\;\sqrt{2ax - x^2}}.\nonumber
\end{equation}
So, if $f=0$ one gets $v^2=h^2-2px$ and one can replace $v^2$ in the expression $\int \frac{fv^2\times\; adx)}{g^2\sqrt{2ax - x^2}}$
by its approximate value $h^2-2px$ as in the preceeding problem. This will give
\begin{eqnarray}
 v^2 &=& h^2-2px -2 \int\frac{fh^2\times\; adx}{g^2\;\sqrt{2ax - x^2}}+2\int\frac{f\times\;2paxdx}{g^2\;\sqrt{2ax - x^2}}\nonumber\\
 &=& h^2-2px -\frac{2fh^2}{g^2} \times\;AM + \frac{2f\times\; 2pa}{g^2}\times\;(AM-MP).\nonumber
\end{eqnarray}
Let $AN$ be the height the pendulum would have reached in vacuum. One has $h^2 = 2p\times\; AN$ and
\begin{equation}
v^2= 2p\times\;PN -\frac{2f\times\;2p\times\;AN\times\;AM}{g^2} +\frac{2f\times\;2pa}{g^2}\times\;(-MP+AM)
\end{equation}
\vskip 0.5cm
\begin{center}
 {\bf Corollary I.}
\end{center}
Hence (fig. 5) when the body arrives at point $c$ such that $Nn= \frac{2f\times\; AN\times\; Ac}{g^2} + \frac{2f\times\;a\times\;(nc-Ac)}{g^2}$
the velocity $v$ will be equal to $0$.

\vskip 0.5cm
\begin{center}
 {\bf Corollary II.}
\end{center}
Since $nc$ and $Ac$ differ little from $NC$ and $AC$, it follows that to find point $c$ where the body stops, or the height $n$ it reaches,
one has to take $Nn=\frac{2f\times\; AN\times\; AC}{g^2} + \frac{2fa\times\;(NC-AC)}{g^2}$.
\vskip 0.5cm
\begin{center}
 {\bf Corollary III.}
\end{center}
If arc $AC$ comprises just a few degrees, $AC$ will be nearly equal $AN$ and one will have $Nn=\frac{2f\times\;AN\times\;AC}{g^2}$ approximately.
\vskip 0.5cm
\begin{center}
 {\bf Corollary IV.}
\end{center}
If a pendulum (fig. 6) descends from $B$, its velocity at $A$, which I called $h$, will be equal (Corol. II, Prop. I) to that it would
have if falling in vacuum from height $An = b - \frac{2fb\times\;BA}{g^2} -\frac{2fa\times\;(BN-BA)}{g^2}$ and it will ascend to height $A\nu$
(Corol. II,
Prop. II) is equal to $An-\frac{2f\times\; An\times\; Ac}{g^2} + \frac{2fa\times\;(nc-Ac)}{g^2}$. And since $nc$ and $Ac$ differ little from $BN$ and
$BA$, we have $A\nu = b-\frac{4fb\times\; BA}{g^2} + \frac{4fa\times\;(BN-BA)}{g^2}$.
\vskip 0.5cm
\begin{center}
 {\bf Corollary V.}
\end{center}
If arc $AB$ comprises but a few degrees, we have $A\nu = b - \frac{4fb\times\; BA}{g^2} = AN\times\frac{(1-4f\times\;BA)}{g^2}$. Or, under the
same assumption, the arcs $AC$ and $Ak$ are to each other approximately as the roots of the abscissae $AN$, $A\nu$. For, in the circle,
the chords are to each other as the roots of the abscissae; or, the arcs can be replaced by the chords. Thus
\begin{equation}
Ck= AC\;\frac{\sqrt{AN}-\sqrt{A\nu}}{\sqrt{AN}}.\nonumber
\end{equation}
or~\footnote{There is a misprint here. The correct form should be $\sqrt{AN\biggl(1-\frac{4f\times\; BA}{g^2}\biggr)}$. [N. of T.]}
\begin{equation}
 \sqrt{A\nu}=\sqrt{AN\;\frac{(1-4f\times\; BA)}{g^2}} = \sqrt{AN}\sqrt{\biggl( 1-\frac{4f\times\; BA}{g^2}\biggr)}.\nonumber
\end{equation}
Since $4f\times\;BA/g^2$ is very small compared to $1$, one may replace $\sqrt{1-\frac{4f\times\; BA}{g^2}}$
by $1-\frac{2f\times BA}{g^2}$, because they are nearly equal and one also knows that if $\alpha$ is a
small fraction, $\sqrt{1-\alpha}$ is approximately equal to $1-\alpha/2$. Thus
\begin{equation}
Ck = AC\times\;\frac{2f\;BA}{g^2}= \frac{2f\;BA^2}{g^2}.
\end{equation}
Hence the difference $Ck$ between the arc $BA$ of descent and the arc $Ak$ of the ascent is as the square of the arc $AB$.
\vskip 0.5cm
\begin{center}
 {\bf Corollary VI.}
\end{center}
Hence (fig. 7) if one knows the arc $BAC$ that a pendulum describes when falling from point $B$, one can easily find the arc
$bAk$ that it will describe when falling from point $b$: it suffices to find $Ak$, which one gets by making
$(BA-AC):(bA-Ak)= BA^2:bA^2$.
\vskip 0.5cm
\begin{center}
 {\bf Corollary VII.}
 \end{center}
Hence it follows that (fig. 6) if a pendulum describes the arc $BA$ in air, one may find its velocity at point $A$ by dividing the line $N\nu$
in two equal parts at point $n$. Then this velocity (Corol. III, Prop. I) is very close to that one would get when falling in 
vacuum from the height $b-\frac{2f\times\;BA}{g^2} = b-\frac{N\nu}{2}$.
\vskip 0.5cm
\begin{center}
 {\bf Corollary VIII.}
\end{center}
One has $AC^2:Ac^2 = An:An$. That is $AC^2:(AC^2-2Cc\times AC) = AN: (AN-Nn)$. Therefore $Nn=\frac{2Cc\times AC\times AN}{AC^2}=
\frac{2Cc\times AN}{AC}$.
For the same reason we have $N\nu= \frac{2Ck\times\;AN}{AC}$. Thus $Ck:Cc = N\nu:Nn$. Therefore $c$ is the middle point of the arcs
$Ck$. Thus, instead of dividing $N\nu$ in two equal parts, we can divide $Ck$ in two equal parts to obtain the arc $Ac$ that
body $A$, on ascencion, would have traversed in vacuum.
\vskip 0.5cm
\begin{center}
 {\bf Corollary IX.}
\end{center}
If pendulum $A$ is a small sphere, the resistance $f$, all other things being equal, is inversely proportional to the
 diameter of this sphere and its density. But the resistance caused by air on two spheres of different diameters goes
 as the surface or the square of the diameter, and this resistance has to be divided by the mass, which is the density
 multiplied by the third power of the diameter. From this it follows that the arc $Ck$, all other things being equal, is like $AB^2$ divided
 by the product of the diameter of the sphere and its density.

 It is up to you, M***, to see if I can now make use of the propositions, since one wants to determine the changes in the
 movement caused by the resistance of air in pendulums used to study the collision of bodies. You will notice, without
 difficulty, that corollaries VI, VII and VIII will give you the velocity that two pendulums would have or receive in the lowest
 point where they supposedely collide.
 
Mr Newton, as you may well know, did not believe in neglecting this resistance. He talked about the collision of bodies
in the first book of his {\it Principia}, and seems to have made $Ck$ proportional not to the square of the arc traversed,
as we found it to be, or as you would suppose, since this was the place in his work that kept you from advancing, but to
the arc solely: and this is what left for me to show you. To this effect, let me transcribe his text, to which I will 
add the comments that I find in the papers that the Reverend Fathers Jacquier and Le Sueur condemned to oblivion, preventing
from their excellent Commentaries, that which I meditated upon.

\begin{center}
 {\bf Newton's Text}
\end{center}
`Let, says Newton ({\it Princip. Mathem. pag. 50, see fig. 8})\footnote{\it Pendeant corpora A, B filis parallelis AC, BD a centris C, D. His centris et intervallis describantur
semicirculi EAF, GBH radijs CA, DB bisecti. Trahatur corpus A ad arcus
EAF punctum quodvis R, et (subducto corpore B)
demittatur inde, redeatq; post unam oscillationem
ad punctum V . Est RV retardatio ex resistentia
aeris. Hujus RV ﬁat ST pars quarta sita in medio, et haec exhibebit retardationem in descensu ab S ad A quam proxime.
Restituatur corpus B in locum
suum. Cadat corpus A de puncto S, et velocitas ejus in loco reﬂexionis A sine
errore sensibili tanta erit, ac si in vacuo cecidisset de loco T. Exponatur igitur haec velocitas per chordam arcus T A;
nam velocitatem Penduli in puncto
inﬁmo esse ut chorda arcus quem cadendo descripsit, Propositio est Geometris
notissima. Post reﬂexionem perveniat corpus A ad locum s, et corpus B ad
locum K. Tollatur corpus B et inveniatur locus v, a quo si corpus A demittatur,
et post unam oscillationem redeat ad locum r, sit st pars quarta ipsius rv sita
in medio, ita videlicet ut rs et tv aequentur; et per chordam arcus tA exponatur velocitas quam corpus A proxime
post reﬂexionem habuit in loco A. Nam t erit locus ille verus et correctus, ad
quem corpus A, sublata aeris resistentia, ascendere debuisset. Simili methodo
corrigendus erit locus k, ad quem corpus B ascendit, et inveniendus locus l, ad
quem corpus illud ascendere debuisset in vacuo. Hoc pacto experiri licet omnia
perinde ac si in vacuo constituti essemus. Tandem ducendum erit corpus A in
chordam arcus T A , quae velocitatem ejus exhibet, ut habeatur motus ejus in
loco A proxime ante reﬂexionem; deinde in chordam arcus tA ut habeatur motus ejus in loco A proxime
post reﬂexionem. Et simili
methodo ubi corpora duo simul demittuntur de locis diversis, inveniendi sunt
motus utriusq; tam ante, quam post reﬂexionem; et tum demum conferendi sunt
motus inter se et colligendi effectus reﬂexionis. Hoc modo in Pendulis pedum
decem rem tentando, idque in corporibus tam inaequalibus quam aequalibus, et
faciendo ut corpora de intervallis amplissimis, puta pedum octo, duodecim vel
sexdecim concurrerent, reperi semper sine errore trium digitorum in mensuris,
ubi corpora sibi mutuo occurrebant, aequales esse mutationes motuum corporibus
in partes contrarias illatae, atque ideo actionem et reactionem semper esse aequales, etc. [Diderot].}
two spherical bodies $A$ and $B$ be suspended from the points
$C$ and $D$ through two equal and parallel lines $AC$ and $BD$ such that these lengths describe two semicircles $EAF$, $GBH$,
divided in two equal parts by the radii $CA$, $CB$. Move body $A$ to any $R$ on the arc $EAF$. Remove $B$ and let $A$ fall: if,
after one oscillation, it returns to point $V$, [then] $RV$ will express the retardation caused by the resistance of air. Take
$ST$ equal to the fourth part of $RV$ and place it in the middle such that $RS$ equals $TV$, that is $RS$ is to $ST$ as $3$ is to
$2$; $RS$ will express very closely the retardation after the descent from $S$ to $A$. Replace the body you have removed. Let $A$ fall
from $S$. Its velocity at the reflection point $A$ will be, without appreciable error, the same velocity it would have if it
had fallen from point $T$. Therefore its velocity will be given by the chord $TA$, as all notable geometers know that the velocity of a pendulum at its lowest
point of the arc it describes is like the chord of this arc. If the body $A$, after the collision, returns to point $S$ and the body
$B$ to point $K$, remove $B$ and find the point $u$ from where $A$, after falling, would return to $r$ such that $st$ is the fourth
part of $ru$ and $sr$ is equal to $tu$. The chord $tA$ will express the velocity $A$ will have at point $A$ after its reflection, since
$t$ is the real and correct place to which $A$ would return in the absence of air resistance. The place $K$ to which body $B$ returns
should be corrected using the same method, finding the point $l$ it would reach in vacuum. This is how one does the experiments as if
in vacuum. Finally one has, so to say, to multiply body $A$ by the chord $TA$ which expresses the velocity, to obtain its movement
at point $A$ immediately before the collision, and by the chord $tA$ to have it right after the collision. One has to search, using
the same method, the quantities of movement before and after collision of two bodies which were let go at the same time from 
two different points, and find, by comparing its movements, the effect of the collision. This is how I performed my experiments
with pendulums 10 feet long, with equal as well as with unequal bodies, which I let fall from afar through distances, for example, of $8$, $12$
and $16$ feet. I found, without having erred in the measured quantities by three inches, that the changes caused by direct collision
in the direction contrary to the movement of the bodies were equal and consequently, action is always equal to reaction, etc.'

\vskip 0.5cm
\noindent
\begin{center}
 {\bf Clarifications}
\end{center}
Here is Newton's text and now the clarifications that I promised to give you. If a body falls from $R$ to $A$ (fig. 9) in a non
resistant medium, its veolcity is, as we know, equal to that it gets when falling from a height equal to that of $RA$. But, as the medium
offers resistance, one may suppose that the velocity [in this case] will be equal to the one it would acquire in a non-resistant medium
when falling through an arc $rA<RA$.

Having reached $A$, if there is no resistance on the branch $AM$, the body will ascend by an arc $A\rho=Ar$; but resistance makes it
ascend no further than $N$. From $N$ it returns to $A$, where we may suppose it has a velocity equal to the one it would have acquired
if it had fallen through the arc $nA<NA$ in a non-resistant medium. And, instead of ascending to $Ay=An$, the resistance of the medium
does not let it ascend beyond $V$.

Put this way, the arc $RV$ expresses the retardations produced by the resistance of air in all the retardation I mentioned. But these
oscillations are each one smaller than the other. To have the retardation of any of them in particular, I have to divide the arc $RV$ into
unequal parts; and as these oscillations are in a number of four, the retardation of the first oscillation is larger than the fourth
part of $RV$; and its fourth part, larger than the retardation of the fourth oscillation. But there is a point $S$ such that a fall
through arc $SA$ will have a retardation given 
exactly by the fourth part of  $RV$. 

Let us find this point $S$. To find it, let $RA=1$; $RV=4b$; $SA=x$. If we assume the retardations are proportional to the traversed arcs,
one will have the retardation $Rr$ of the arc $RA$ traversed is equal to $\frac{b}{x}$. And $A\rho$, the second arc, is equal to $AR = RA-Rr= 1-\frac{b}{x}$. The same way
$\rho N$, the retardation of arc $A\rho$, is $(1-\frac{b}{x})\times\;\frac{b}{x} = \frac{b}{x} - \frac{b^2}{x^2}$. Hence $AN$, the third arc,
is $A\rho -\rho nN= 1-\frac{2b}{x} +\frac{b^2}{x^2}$. And the retardation $Nn$ of the arc
$AN$ is $\biggl(1-\frac{2b}{x} +\frac{b^2}{x^2}\biggr)\times\frac{b}{x} = \frac{b}{x}-\frac{2b^2}{x^2}+\frac{b^3}{x^3}$. Hence
$Ay=An=AN-Nn$, the fourth arc, is equal to $1-\frac{3b}{x}+\frac{3b^2}{x^2}+\frac{3b^3}{x^3}-\frac{b^4}{x^4}$. Thus, the retardation $Vy$ of the fourth arc 
is $\frac{b}{x} -\frac{3b^3}{x^2} + \frac{3b^3}{x^3} -\frac{b^4}{x^4}$. One thus has
\begin{itemize}
 \item[] $Rr$, retardation of the first arc, is equal to $\frac{b}{x}$.
 \item[] $\rho N$, retardation of the second, is equal to $\frac{b}{x} - \frac{b^2}{x^2}$.
 \item[] $Nn$, retardation of the third, is equal to $\frac{b}{x}-\frac{2b^2}{x^2}+\frac{b^3}{x^3}$.
 \item[] $Vy$, retardation of the fourth, is equal to $\frac{b}{x} -\frac{3b^3}{x^2} + \frac{3b^3}{x^3} -\frac{b^4}{x^4}4$.
\end{itemize}
And since $Rr+\rho N+Nn+Vy= 4b$, we have the equation $\frac{4b}{x} -\frac{6b^2}{x^2}+\frac{4b^3}{x^3} -\frac{b^4}{x^4}=4b$, or
$x^4 - x^3+\frac{3bx^2}{2}-\frac{b^2}{x^2} +\frac{b^3}{4}=0$, whose approximate solution will give us the value of $x$.

To find it, we drop the last two terms $-\frac{b^2}{x^2} +\frac{b^3}{4}$ which are much smaller than the other terms since $b$ is
very small. One is left with $x^4 - x^3+\frac{3bx^2}{2}=0$, or $x^2 - x+\frac{3b}{2}=0$, an equation whose root is
$x=\frac{1}{2}+\sqrt{\frac{1}{4}-\frac{3b}{2}}$. However, $\sqrt{\frac{1}{4}-\frac{3b}{2}}$ is very close to $\frac{1}{2}-\frac{3b}{2}$
from which [we have that]
$x$ is very close to $\frac{1}{2}+\frac{1}{2}-\frac{3b}{2}=1-\frac{3b}{2}$.
\vskip 0.5cm
\begin{center}
 {\it Remarks on this approximation}
\end{center}
\begin{itemize}
\item[{\it 1st.}] Notice that $-b^2x^2+\frac{b^3}{4}<0$, since $x>b$, from which it follows that $x^4 - x^3+\frac{3bx^2}{2}=0$. Thus
$x>\frac{1}{2}+\sqrt{\frac{1}{4}-\frac{3b}{2}}$.
But $\frac{1}{2}-\frac{3b}{2}$ is little larger than $\sqrt{\frac{1}{4}-\frac{3b}{2}}$, so when replacing  $\sqrt{\frac{1}{4}-\frac{3b}{2}}$ by
$\frac{1}{2}-\frac{3b}{2}$ one is giving back to $x$
a little of what one took from it. From which it follows that this approximation is as simple and accurate as we could wish,
given the assumption that the retardations are as the arcs and not the square of the arcs.\\

\item[{\it 2nd.}] The retardations $\frac{b}{x}$; $\frac{b}{x}-\frac{b^2}{x^2}$, etc.  are in geometrical progression.\\

\item[{\it 3rd.}] One may solve the equation $\frac{4b}{x} -\frac{6b^2}{x^2}+\frac{4b^3}{x^3} -\frac{b^4}{x^4}=4b$ exactly if one makes the approximation
$1-\frac{4b}{x} +\frac{6b^2}{x^2}-\frac{4b^3}{x^3} +\frac{b^4}{x^4}=1-4b$. Thus $1-\frac{b}{x} = \sqrt[4]{1-4b}$ or $x=\frac{b}{1-\sqrt[4]{1-4b}}$.\\

\item[{\it 4th.}] That, to find the place $V$, we have $st:tu = 2:3$ and that $tu=sr$. From which it follows that
$su:sr = 5:3$. Let $As=1$, $sr=x$. We have $Au= 1+\frac{5x}{4}$; $Ar= 1-x$. Or $Au$ is to $Ar$ almost as $AV:AR$. So, if we
make $AV:AR = m:n$, we will have $m:n = (1=\frac{5x}{3}):(1-x)$. From this $n+\frac{5xn}{3}=m-mx$ and 
$\frac{m-n}{m+5n/3} = \biggl(3\times \frac{m-n}{3m+5n}\biggr)\times\; As$ since we assumed $As=1$.
\end{itemize}
One may now determine the point $V$ through experiments, letting a pendulum fall from point $V$ until it returns to a point $r$,
where the distance $sr$ [from $r$] to $s$ is such that it is equal to $su\times\frac{3}{5}$ or, one may simply take $st= \frac{As}{AS}\times\; ST$.

Here is, to me very well clarified it seems, the entire passage of Newton on the retardation of pendulums caused
by the resistance of air. From it, there seems to follow that this author assumed the retardations [proportional] to
the arcs while, according to the preceeding propositions,  we found it to be as the squares of the arcs. You may object,
undoubtedly, that Newton has the experiments on his side; and that with this hypothesis he found action to be always
equal to reaction\footnote{Ut si corpus A incidebat
in corpus B quiescens cum novem partibus motus, et amissis septem partibus pergebat post reflexionem cum duabus; corpus B resiliebat
cum partibus istis septem. Si corpora obviam ibant, A cum duodecim partibus et $B$ cum sex, et redibat A cum duabus; redibat B cum octo,
facta etc. [Diderot].}; and that, for example,
if body $A$ with $9$ degrees of movement, after colliding with $B$ initially at rest, kept moving with $2$ while $B$ departed with $7$; that
if bodies collided coming from opposite directions, $A$ with $12$ degrees, $B$ with $6$ and that $A$ reflected with $2$
$B$ reflected with $8$, etc.

Being advised to never doubt the exactitude and good faith of Newton, I would nonetheless like to remind you that
this did not prevent his experiments on colors from being repeated. Why would one not do this in this particular case, where
the author started with an hypothesis which calculations clearly contradict and where it is even easier to make
a mistake since the velocities are represented by quantities whose differences are very small, namely, the chords of the
arcs traversed before and after retardations? 

If you think this is not enough, since it is Newton's great name [after all], I am vexed; as for me, I cannot agree with
him. I have for Newton all deference one may accord to the unique men of his kin and tend strongly to the belief that he has
truth at his side. But, even so, it is better to be sure of if. I invite therefore all those who like the good physics to
restart their experiments and tell us if the retardations are as those that Newton seems to have assumed, proportional to the
arcs traversed; or those that the calculations give us, proportional to the square of these arcs.

\end{document}